\documentclass[alpha-refs]{wiley-article}

\usepackage{color}
\usepackage{ulem}
\usepackage{natbib}
\usepackage{textcomp}
\usepackage[export]{adjustbox}
\usepackage{xr}
\usepackage{caption}
\makeatletter
\newcommand*{\addFileDependency}[1]{
  \typeout{(#1)}
  \@addtofilelist{#1}
  \IfFileExists{#1}{}{\typeout{No file #1.}}
}
\makeatother

\newcommand*{\myexternaldocument}[1]{%
    \externaldocument{#1}%
    \addFileDependency{#1.tex}%
    \addFileDependency{#1.aux}%
}

\myexternaldocument{supplement}

\usepackage{siunitx}

\papertype{Research Article}

\title{The link between North Atlantic tropical cyclones and ENSO in seasonal forecasts}

\abbrevs{TC, tropical cyclone; ENSO, El Niño-Southern Oscillation; JASO, July-August-September-October; MDR (-A,-C), Main Development Region (-Atlantic, -Caribbean); GOM, Gulf of Mexico}

\author[1]{Robert Doane-Solomon}
\author[1]{Daniel J. Befort}
\author[2]{Joanne Camp}
\author[3,4]{Kevin Hodges}
\author[5,6]{Antje Weisheimer}

\affil[1]{Atmospheric, Oceanic and Planetary Physics, Department of Physics, University of Oxford, Oxford, United Kingdom}
\affil[2]{Met Office Hadley Centre, Exeter, United Kingdom}
\affil[3]{Department of Meteorology, University of Reading, Reading, United Kingdom}
\affil[4]{National Centre for Atmospheric Science, University of Reading, Reading, RG6 6ET, UK}
\affil[5]{National Centre for Atmospheric Science, Atmospheric, Oceanic and Planetary Physics, Department of Physics, University of Oxford, Oxford, United Kingdom}
\affil[6]{European Centre for Medium-Range Weather Forecasts (ECMWF), Reading, UK}

\corraddress{Robert Doane-Solomon, Pembroke College, University of Oxford, Oxford, OX1 1DW, United Kingdom}
\corremail{robert.doane-solomon@pmb.ox.ac.uk}

\fundinginfo{RDS was funded by the Met Office Academic Partnership (MOAP). DJB and AW were supported by the EUCP project funded by the European Commission’s Horizon 2020 programme, Grant Agreement number 776613. KH was funded as part of the NERC National Centre for Atmospheric Science.}

\runningauthor{Robert Doane-Solomon et al.}

\begin{document}

\maketitle

\begin{abstract}
This study assesses the ability of six European seasonal forecast models to simulate the observed teleconnection between ENSO and tropical cyclones (TCs) over the North Atlantic. While the models generally capture the basin-wide observed link, its magnitude is overestimated in all forecast models compared to reanalysis. Furthermore, the ENSO - TC relationship in the Caribbean is poorly simulated. It is shown that incorrect forecasting of wind shear appears to affect the representation of the teleconnection in some models, however it is not a completely sufficient explanation for the overestimation of the link.

\end{abstract}

\newpage
\section{Introduction}
Tropical cyclones (TCs) cause significant annual losses and casualties, with North Atlantic tropical cyclones being the most destructive natural disasters in the United States \citep{Grinsted23942}. Several variables affect the interannual variability of TC numbers and their location, such as the El Niño–Southern Oscillation (ENSO) \citep{xie2005climatology, ENSOcyclonesbook, jaramillo2021combined} and tropical Atlantic sea surface temperatures \citep{KlotzbachPhilip2019STCF}. Years exhibiting La Niña conditions since 2010 have been associated with record-breaking tropical cyclone activity in the North Atlantic in 2020 and an extremely active season in 2017 \citep{klotzbach2022hyperactive, lim2018roles}. Indeed, ENSO is thought to be the most important single factor in seasonal tropical cyclone variability \citep{camargobook, YangSong2018ENOa, chu2004enso, gray1984atlantic}. Global shifts in atmospheric and oceanic weather patterns occur in response to ENSO, due to the strengthening/weakening of the Walker and Hadley circulations. Over the North Atlantic, ENSO impacts include changes in sea surface temperatures \citep{goldberg}, tropospheric vorticity \citep{camargo2007use}, vertical wind shear and humidity \citep{ImpactonAtlanticBasin}, all of which can have an impact on TCs. In an El Niño year wind shear usually increases in the Main Development Region of the tropical Atlantic (MDR - the area between 10-20N \& 20-85W), and vice versa during La Niña years \citep{camargobook}. TCs can be thought of as heat engines transferring heat from warm tropical ocean waters to the cold upper atmosphere - wind shear suppresses a TC by dissipating warm air from the top of the storm and inducing asymmetries in its circulation, making the heat engine less efficient \citep{frank2001effects, willoughby1999hurricane}. Consequently, TC activity usually decreases in the North Atlantic during El Niño, whereas a La Niña event usually enhances TC activity \citep{BrownMichaelE2006HaTP, ImpactonAtlanticBasin}. Due to their socio-economic impacts, skillful seasonal predictions of TC activity are crucial for disaster planning for nations bordering the Atlantic, Caribbean and Gulf of Mexico. To obtain skillful seasonal forecasts of TC activity it is important to correctly represent the observed relationship between tropical Pacific sea surface temperatures and those factors impacting North Atlantic TCs, e.g. wind shear and local SSTs.
\newline
\indent Many operational forecasting centers produce seasonal forecasts with approximately 6 month lead times from which predictions of TC activity can be determined. Previous studies show that the interannual variability of NA TCs can be predicted to some extent using such seasonal forecasts. \cite{vitart2007dynamically} studied 3 seasonal forecast models from the European Seasonal to Interannual Prediction (EUROSIP) system, finding a high degree of correlation between predicted and actual NA TC numbers per season. More recently a study analysed the ability of 6 state-of-the-art operational European seasonal forecast models in simulating interannual TC variability over the NA and Western North Pacific basins during 1993 to 2014 \citep{befort2022seasonal}. They found that these models are skillful in simulating TC variability over the NA basin and generally poor over the Western North Pacific, despite significant climatological differences between the models. However, \cite{befort2022seasonal} did not investigate the ability of the models to simulate the observed TC-ENSO teleconnection in either of these basins. ENSO forecast accuracy is a crucial factor for the correct prediction of seasonal TC variations \citep{camargo12007seasonal}. Analysis of the UK Met Office Glosea5-GA3 model over the 1992-2013 period has shown a statistically significant reduction in NA TC numbers during periods of El Niño activity \citep{CampJ2015Sfot}. Additionally, \cite{vitart2001seasonal} analysed the teleconnection between ENSO and NA TCs in an early version of ECMWF's seasonal forecast system. This correctly showed fewer storms in El Niño \& more in La Niña, albeit with a small sample size from 1991-1999.  

\indent Here, we expand on the \cite{befort2022seasonal} study by analysing the predictability of the TC-ENSO teleconnection in the North Atlantic for the same seasonal forecast models. This paper is structured as follows: first we assess the relationship between central equatorial Pacific SSTs and TC numbers during the active July to October (JASO) season in the different seasonal forecast models, and compare them to reanalysis and best track observations. Next, we analyse to what extent the errors in the simulated ENSO-TC relationship can be explained by local environmental variables over the tropical NA. This involves investigating the simulated link between ENSO and the interannual variability of vertical wind shear \& humidity, and how differences between the simulated relationships may contribute to errors in the overall simulated interannual variability of TCs. We conclude by summarising the results and discussing the limitations of the methods and data used.

\section{Methods and Data}
In this study, data from 6 different dynamical seasonal forecast models are used, which are the same models as in \cite{befort2022seasonal}. They are from the following six European models which contribute to the Copernicus Climate Change Service (C3S) multi-model seasonal forecasting system.: UK Met Office GloSea5-GC2 (UKMO) \citep{UKMO}, German Weather Service GCFS2.0 (DWD) \citep{DWD}, Euro-Mediterranean Centre on Climate Change SPS3 (CMCC) \citep{CMCC}, European Centre for Medium-Range Weather Forecasts SEAS5 (ECMWF) \citep{ECMWF}, Météo-France System 5 (MF5) and Météo-France System 6 (MF6) \citep{MF}. Details on the resolution and ensemble size of each of these models are available in Table \ref{model table} in the supplementary material. Forecasts initialised at the beginning of June during the hindcast period 1993 and 2014 are used, which results in a total of 22 TC seasons. The European Centre for Medium-Range Weather Forecasts (ECMWF) 5th generation reanalysis (ERA5) \citep{ERA5} is used as a reference for the ENSO Index and TC numbers, as well as specific humidity and wind shear. This reanalysis covers the globe using a 31km grid, similar to that of the highest resolution forecast models in this study (Table \ref{model table}). The same tracking and detection methods are used to identify TCs in ERA5 as in the 6 seasonal forecast models. Best Track data from the International Best Track Archive for Climate Stewardship (IBTrACS) \citep{IBTrACS} is also used an additional reference for TC counts. IBTrACS is a dataset of historical storms' location and intensity, interpolated to 3-hourly intervals with 0.1\textdegree resolution. We sub-sample this data to 12-hourly intervals to be consistent with the seasonal forecasting systems.

\indent The tropical cyclones in ERA5 and the seasonal forecast models are tracked using the methods described in \cite{HodgesKevin2017HWAT}. Potential tropical cyclones are first identified using the vertical average of vorticity between 850 and 700hPa, spectrally filtered to T63 with the large scale background removed, and a threshold of $5 \times 10\textsuperscript{-6} s \textsuperscript{-1}$. The potential cyclones are then used to track the motion of the storms. By minimizing a cost function subject to adaptive constraints, smooth storm trajectories are produced \citep{hodges1994general, hodges1995feature, hodges1999adaptive}. Tropical cyclones are then identified by using the following criteria, the same as in \citep{befort2022seasonal, HodgesKevin2017HWAT}:
\nobreak
\begin{enumerate}
\item The T63 relative vorticity at 850hPa must be at least $6 \times 10\textsuperscript{-5} s \textsuperscript{-1}$
\item The T63 vorticity difference between 850 and 200 hPa must be at least $6 \times 10\textsuperscript{-5} s \textsuperscript{-1}$
\item There must be a T63 vorticity centre at each level between 850hPa and 200hPa for which there is data
\item The above criteria must apply for a minimum of 2 consecutive timesteps
\item Tracks must start between 0N-30N
\end{enumerate}

\noindent Gridded track density statistics are calculated using the spherical kernel estimators described in \cite{hodges1996spherical}. The JASO (July-August-September-October) period is used for all analysis, being the most active season over the NA basin \citep{befort2022seasonal}. The Niño 3.4 region (5S-5N, 170W-120W) is used for the ENSO index in our analysis. The index used here is equivalent to the SST anomaly in this region, using 1993-2014 as the baseline (Niño 3.4 Index). Study of the North Atlantic is performed basin-wide, with particular interest paid to the following 3 regions: Main Development Region - Atlantic (MDR-A) (10-20N, 20-60W),  Main Development Region - Caribbean (MDR-C) (10-20N, 60-85W) and Gulf of Mexico (GOM) (20-30N, 100-80W) \citep{ercolani2015intense, burn2015atlantic}. These 3 regions have high tropical cyclone track densities and/or landfall potential \citep{weinkle2012historical}. In all maps of the basin shown, areas are masked where the tropical cyclone track density in ERA5 is less than half the basin-wide average. The basin is defined as the regions (10-20N, 20-85W), (20-30N, 20-100W) \&  (30-40N, 20-80W). This is to help the analysis by removing areas with low tropical cyclone density.

\indent This study focuses on the TC - ENSO teleconnection. Connectors and relationships like TC - wind shear, wind shear - ENSO, TC - humidity \& humidity - ENSO are also analysed. Here, vertical wind shear (WS) is calculated using the following formula:

\begin{equation*}
  \text{VWS}\ = \sqrt{(u_{200}-u_{850})^2 + (v_{200}-v_{850})^2}
\end{equation*}

\noindent where \(u\) and \(v\) refer to zonal and meridional winds, and the subscript refers to the pressure level. If tropical cyclone numbers are being compared, IBTrACS is also used as a reference. The forecast model performance is assessed by comparing linear correlation and regression coefficients from the forecasts with those from ERA5 and IBTrACS. Spatial statistics are generated by calculating temporal correlation coefficients between two variables at each grid point (e.g. wind shear and TC numbers), or between one variable and the ENSO Index. Bootstrapping is used to provide an estimate of the uncertainty of regressions and correlations. This is done by resampling the target data 1000 times with replacement. For the purposes of this study, the values at the 10th and 90th percentiles of a bootstrapped distribution were used to provide estimates of upper and lower bounds for that value.
\newline
\indent TC track composites of El Niño years and La Niña years (as predicted by that model) were also used to investigate any nonlinear behaviour. For a given model, El Niño years are defined as when that model had a JASO ENSO Index $> 0.5 $\textdegree C, or $< -0.5$\textdegree C for La Niña years. These years are shown in Supplementary Table \ref{years table} below. TC track density anomalies for these years are then standardised. This is done by dividing the anomaly by the standard deviation of the number of tracks that cross that point at the respective grid cell.

\section{Results}
\subsection {Impact of ENSO on TC numbers}
\begin{figure}[!htb]
    \centering
    \includegraphics[width=0.6\textwidth]{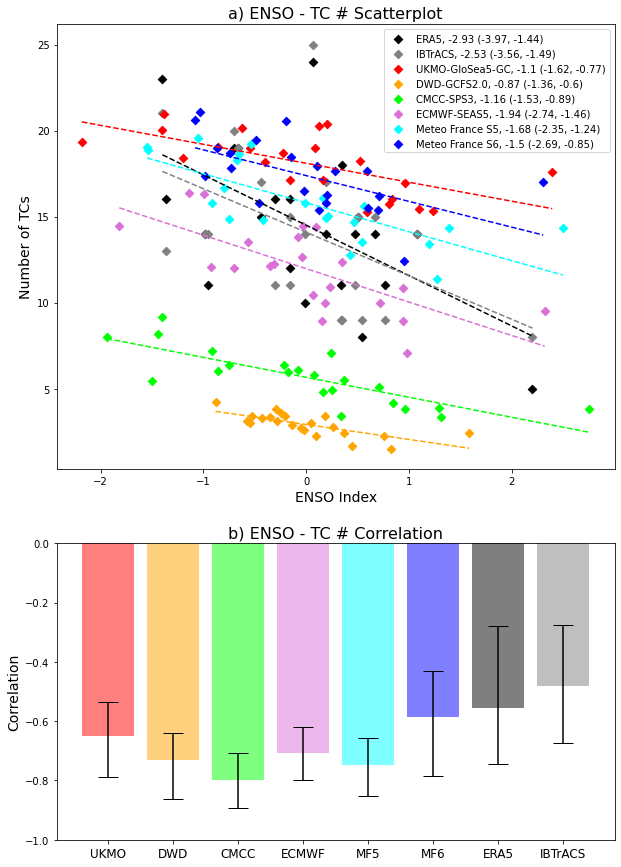}
    \caption{a): Scatterplot of number of TCs in the North Atlantic basin for JASO from 1993-2014 vs NINO 3.4 SST Index for IBTracks (grey), reanalysis (black) and the seasonal forecast models (colours). Values in the legend are the regression coefficient between the two variables, and the values in brackets are the 10th \& 90th percentiles of the bootstrapped regression distribution. b): Bar plot of the correlation between the number of TCs in the North Atlantic basin for JASO from 1993-2014 and NINO 3.4 SST index. Error bars are the 10th \& 90th percentiles of the bootstrapped correlation distribution.}
    \label{tc-enso scatter and bar}
\end{figure}

We initially investigate the  relationship between ENSO and North Atlantic TCs on a basin-wide scale (Figure \ref{tc-enso scatter and bar}). As shown in Figure \ref{tc-enso scatter and bar}a, a negative relationship between the ENSO Index and TC numbers is found when using IBTrACS or tracks identified in ERA5, which is in line with previous studies \citep{BrownMichaelE2006HaTP, ImpactonAtlanticBasin}. All 6 forecast models capture this negative relationship between ENSO and TC numbers, despite the fact that some models (DWD, CMCC) have significant negative biases in TC numbers. The magnitude of the regression coefficient (shown in the legend of Fig. \ref{tc-enso scatter and bar}a)  between ENSO and TC numbers, ranges from approximately -2.5 to -3 in IBTrACS and ERA5 and approximately -1 to -2 in the models or similar. Figure \ref{tc-enso scatter and bar}b shows the correlation coefficient, with the error bars corresponding to the correlation coefficients at the 10th \& 90th percentiles of the bootstrapped distribution. We see that for all the seasonal forecast models, the central correlation value is larger in magnitude than for ERA5 and IBTrACS, however the regression coefficients are all smaller in magnitude. This is likely due to the interannual variability of the ensemble mean number of storms predicted by the seasonal forecasts being much lower than in ERA5 and IBTrACS. However, the central correlation value of all the models are consistent with the uncertainty range of ERA5, except for CMCC whose which lies just outside. The similarity of the correlation and regression values (and their respective uncertainty) between ERA5 and IBTrACS in Figure \ref{tc-enso scatter and bar} motivates the use of ERA5 alone as the comparison in the following plots.

\begin{figure}[!htb]
    \centering
    \includegraphics[width= 0.4\textwidth]{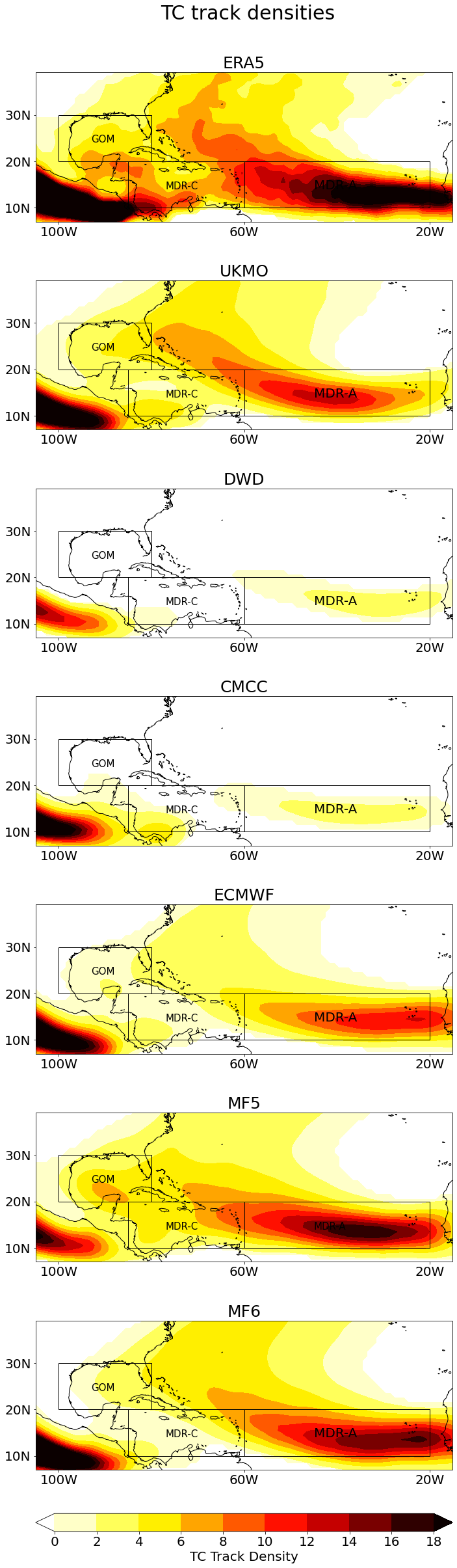}
    \caption{Map of TC track densities from reanalysis and forecast model data. Densities are shown where there is on average $\ge 1$ storm per season.}
    \label{track densities}
\end{figure}

\indent We next analyse the effect of ENSO on the TC spatial pattern across the basin. First, we show average JASO track densities across the basin in Figure \ref{track densities}. The ERA5 data shows the MDR-A (the central Atlantic between 10 \& 20N) is the most active region for TC development. Many storms there track north-westwards and recurve in the open Atlantic, while some move westwards into the MDR-C (the Caribbean). In the western part of MDR-C and the GOM (Gulf of Mexico), an area of enhanced activity is also seen. A similar pattern in MDR-A is seen in the models, however some models (UKMO, MF6) have too large a proportion of storms on a recurving track while others appear slightly too zonal (MF5). As in Figure \ref{tc-enso scatter and bar}a, we see that the DWD and CMCC models produced far fewer storms than ERA5. 

\indent The track density changes in ENSO years (SI Table \ref{years table}) are shown in the composites of standardised track density anomalies for warm and cold events (Figure \ref{composites}). For El Niño years in ERA5 (Fig. \ref{composites}a), we find two main areas of lower track density: one over the MDR-A and recurving northwards, and another in the western MDR-C \& GOM. These two areas are separated by an area of \textit{increased} TC activity in the eastern MDR-C and near Florida. The length of the period used here (1993-2014) does not appear to affect the robustness of the pattern in ERA5, as we find similar results when analysing the period from 1979 to 2017 (Fig. \ref{era5 comparison} in SI). This is consistent with previous studies \citep{climatemodelsenso, jaramillo2021combined, camargo2007use}. All models capture the basin-wide reduction in TC activity during El Niño years, but are unable to capture the finer spatial pattern described above. During El Niño years, no model predicts an increase anywhere in MDR-C. For La Niña years in ERA5 (Fig. \ref{composites}h), we find TC track density anomalies are essentially the opposite of the response during El Niño years. In La Niña, we see an increase in storm count over the MDR-A and in the western MDR-C \& GOM, and a decrease in the eastern MDR-C and near Florida. The models do not capture this pattern, but the basin-wide average response is captured well during La Niña years.

\indent In Fig 1 we saw that the interannual variability between basin-wide TC counts and ENSO is well represented in the models. Here we analyse the spatial structure of this relationship across the Atlantic basin. Figure \ref{TC-ENSO map} shows the correlation between ENSO Index \& TC numbers in different areas of the basin. ERA5 displays a statistically significant negative correlation in MDR-A and the western MDR-C \& GOM regions. This is in line with the basin-wide results in Figure \ref{tc-enso scatter and bar}. Weak negative correlation is seen in the majority of the MDR-C region \& near Florida. Most forecast models in general predict well the strong negative correlation in MDR-A (except for MF6). However, many models show a stronger negative correlation near Florida and in MDR-C, which is not found in ERA5. We also note that many models predict an area of significant negative correlation around 30N, though the size and location of this differs between them. Such a signal is not seen in ERA5. Along with the results from Figure \ref{composites}, this provides further evidence that despite the good overall basin-wide response to different ENSO phases, all analysed forecast models have difficulties representing the details of the spatial pattern of the ENSO - TC teleconnection over the North Atlantic basin.

\subsection{Relationship between ENSO, TC numbers and environmental variables}
To explore what might be causing some of the errors found in the TC-ENSO relationship in the different models, we examine the effect of environmental variables that link the ENSO SST with the atmospheric conditions required for the formation of TCs in the NA basin. Wind shear (WS) and humidity are among these variables heavily influenced by ENSO. For example, during El Niño the Walker circulation shifts eastwards, leading to stronger high-altitude westerly winds within the Hadley cell, and descending dry air over the Caribbean. This in turn suppresses the formation of TCs \citep{camargobook, ImpactonAtlanticBasin}. 
\newline
The correlations between ENSO and wind shear, and between wind shear and TC numbers, are shown in Figure \ref{Shear map}. The ERA5 reanalysis (Fig. \ref{Shear map}a) shows that the ENSO Index is positively correlated with the wind shear across most of the Atlantic basin, except for the wider region around Florida. Wind shear is generally negatively correlated with TC numbers across the basin, as expected \citep{gray1968global}. South of 25N, this is replicated in the models, though with slight shifts in location and intensity. The magnitude of the correlation is overestimated in the MDR-A for all models, except for MF6 which shows a large region with correlations around 0. However, the observed ENSO-WS/WS-TC relationship north of 25N is uniformly poorly simulated by the models - they display a sharp transition in the sign of the correlation coefficient, in contrast to ERA5. 
\newline
\indent As moist air is another crucial element for TC development and is known to be affected by ENSO, we performed the same analysis for 500hPa specific humidity. The results are available in the supplementary material (Figure ref{Humidity map}), however it was judged that errors in forecast models' representation of the ENSO - TC teleconnection was more likely due to wind shear than humidity. This is because in the locations with the largest errors in the ENSO - TC correlation, the ENSO-humidity and humidity - TC correlations are generally well simulated (around 30N and in the MDR-C).

\section{Conclusions and Discussion}
We expand on a recent study by \cite{befort2022seasonal}, analysing the performance of six European seasonal forecast models at predicting the TC-ENSO teleconnection in this region. Basin-wide, it is found that the observed negative correlation between the Niño 3.4 ENSO index and the total number of TC counts over the season is well predicted by all of the models studied. The correct sign of the correlation coefficient is usually forecast correctly in the MDR and GOM; occasionally no signal is seen (as in MF6). However in the eastern Caribbean, the models simulate a statistical relationship that is not present in ERA5. Errors in this region are also seen in the ENSO composite analysis (Figure \ref{composites}). Similar errors are found by \cite{climatemodelsenso} when analysing the response of global climate models to Central Pacific El Niño events (their Figure 6), which they partially attribute to errors in simulating wind shear. However in our analysis, we see no firm evidence that wind shear and humidity drive the spatial differences in the MDR-C between the models and ERA5. However, the magnitude of the correlation is generally too large in the forecasts, especially in between 10 \& 20N: this is due to models consistently overestimating the relationship between local variables and TC numbers.
\newline
\indent The sharp transition in the sign of the correlation coefficient in the forecast models north of 25N may be due to errors simulating the Hadley cell's response to ENSO. It is believed that the northern edge of the Hadley cell usually moves south during El Niño and north during La Niña \citep{wang2004enso, HU2018640, lu2008response}. In theory this would lead to a strong correlation between ENSO and wind shear in the tropics inside the Hadley cell, with a smooth decrease to an area of no correlation on the north side of the boundary. As this is not seen in the models, the sharp decrease may be due them failing to alter the width of the Hadley cell in response to ENSO, keeping the edge location fixed. This incorrect modelling of the edge of the Hadley cell may cause the pattern seen in the simulated correlation between wind shear \& TCs, which is strongly positive in the subtropics despite the physical effect of wind shear causing storms to weaken and dissipate. This is not seen in ERA5 which has a weak negative correlation almost throughout the NA basin. Together this may explain the areas of negative correlation seen in the subtropics in the overall ENSO - TC correlation (Figure \ref{TC-ENSO map}).
\newline
\indent In general, it is found that the forecast models simulate statistical relationships between the environmental variables and TC numbers (as opposed to ENSO and environmental variables) for larger areas of the North Atlantic than can be detected in reanalysis data, especially in the subtropics. One reason for this is the fact that the seasonal forecast models use a coarse horizontal resolution of 50-80 km. Secondly, it is also likely due to the fact that there is only 1 realization for each year in ERA5. However, there are tens of ensemble model runs for each year in the forecasts which are combined into the ensemble mean. So it is likely the area of significant correlations with TC numbers would be larger in the forecasts, as the sample size is much larger. To address this issue we computed the distribution of individual correlations from randomly sampled ensemble members \textit{before} taking the mean. However, the TC locations vary considerably between ensemble members and so significant correlations are only seen where the TCs are identified. So when the mean is taken, the average correlation in any one location is very low, leading to a poor signal to noise ratio.  However, by bootstrapping the data as in figures \ref{tc-enso scatter and bar}, we can still estimate the uncertainty of the correlations and whether the forecast models are consistent with ERA5. 
\newline
\indent This analysis then shows that the basin-wide TC-ENSO teleconnection is well represented by these 6 seasonal forecasts between 1993 and 2014. However, due to the limitations of a small sample size, the uncertainty around the notable spatial and intensity errors is large. Further research including longer hindcasts will allow for more robust results.
\section{Conflicts of Interest}
All authors declare they have no conflicts of interest. 
\section{Acknowledgements} 
We are grateful for the support received from the Met Office Academic Partnership (MOAP). Robert Doane-Solomon received funding from MOAP. Daniel J. Befort and Antje Weisheimer received funding from the European Union under Horizon 2020 (Grant Agreement 776613. Kevin Hodges was funded as part of the NERC National Centre for Atmospheric Science. All reanalysis and seasonal forecast data is available from the C3S website (\url{https://climate.copernicus.eu}). Tropical cyclone track data is available upon request for the seasonal forecast models and reanalysis. IBTrACS data is available at \url{https://www.ncei.noaa.gov/products/international-best-track-archive}.
\newpage
\bibliography{library}

\begin{thebibliography}{42}
\expandafter\ifx\csname natexlab\endcsname\relax\def\natexlab#1{#1}\fi
\expandafter\ifx\csname url\endcsname\relax
  \def\url#1{\texttt{#1}}\fi
\expandafter\ifx\csname urlprefix\endcsname\relax\def\urlprefix{URL: }\fi

\bibitem[{Befort et~al.(2022)Befort, Hodges and
  Weisheimer}]{befort2022seasonal}
Befort, D.~J., Hodges, K.~I. and Weisheimer, A. (2022) Seasonal prediction of
  tropical cyclones over the north atlantic and western north pacific.
\newblock \textit{Journal of Climate}, \textbf{35}, 1385--1397.

\bibitem[{Brown(2006)}]{BrownMichaelE2006HaTP}
Brown, M.~E. (2006) Hurricanes and typhoons: Past, present, and future.
\newblock \textit{Southeastern Geographer}, \textbf{46}, 323--324.

\bibitem[{Burn and Palmer(2015)}]{burn2015atlantic}
Burn, M.~J. and Palmer, S.~E. (2015) Atlantic hurricane activity during the
  last millennium.
\newblock \textit{Scientific reports}, \textbf{5}, 1--11.

\bibitem[{Camargo et~al.(2007)Camargo, Emanuel and Sobel}]{camargo2007use}
Camargo, S.~J., Emanuel, K.~A. and Sobel, A.~H. (2007) Use of a genesis
  potential index to diagnose enso effects on tropical cyclone genesis.
\newblock \textit{Journal of Climate}, \textbf{20}, 4819--4834.

\bibitem[{Camargo et~al.(2010)Camargo, Sobel, Barnston and
  Klotzbach}]{camargobook}
Camargo, S.~J., Sobel, A.~H., Barnston, A.~G. and Klotzbach, P.~J. (2010)
  \textit{The Influence of Natural Climate Variability on Tropical Cyclones,
  and Seasonal Forecasts of Tropical Cyclone Activity}, 325--360.
\newblock World Scientific.
\newblock
  \urlprefix\url{https://www.worldscientific.com/doi/abs/10.1142/9789814293488_0011}.

\bibitem[{Camargo$^1$ et~al.(2007)Camargo$^1$, Barnston, Klotzbach and
  Landsea}]{camargo12007seasonal}
Camargo$^1$, S.~J., Barnston, A.~G., Klotzbach, P.~J. and Landsea, C.~W. (2007)
  Seasonal tropical cyclone forecasts.
\newblock \textit{Wmo Bulletin}, \textbf{56}, 297.

\bibitem[{Camp et~al.(2015)Camp, Roberts, MacLachlan, Wallace, Hermanson,
  Brookshaw, Arribas and Scaife}]{CampJ2015Sfot}
Camp, J., Roberts, M., MacLachlan, C., Wallace, E., Hermanson, L., Brookshaw,
  A., Arribas, A. and Scaife, A.~A. (2015) Seasonal forecasting of tropical
  storms using the met office glosea5 seasonal forecast system.
\newblock \textit{Quarterly journal of the Royal Meteorological Society},
  \textbf{141}, 2206--2219.

\bibitem[{Chu(2004)}]{chu2004enso}
Chu, P.-S. (2004) \textit{ENSO and tropical cyclone activity}.
\newblock Columbia University Press New York.

\bibitem[{Dorel et~al.(2017)Dorel, Ardilouze, Batté, Déqué and
  J.~Guérémy}]{MF}
Dorel, L., Ardilouze, C., Batté, L., Déqué, M. and J.~Guérémy, . (2017)
  Documentation of the 631 météo-france pre-operational seasonal forecasting
  system.

\bibitem[{Ercolani et~al.(2015)Ercolani, Muller, Collins, Savarese and
  Squiccimara}]{ercolani2015intense}
Ercolani, C., Muller, J., Collins, J., Savarese, M. and Squiccimara, L. (2015)
  Intense southwest florida hurricane landfalls over the past 1000 years.
\newblock \textit{Quaternary Science Reviews}, \textbf{126}, 17--25.

\bibitem[{Frank and Ritchie(2001)}]{frank2001effects}
Frank, W.~M. and Ritchie, E.~A. (2001) Effects of vertical wind shear on the
  intensity and structure of numerically simulated hurricanes.
\newblock \textit{Monthly weather review}, \textbf{129}, 2249--2269.

\bibitem[{Fröhlich et~al.(2020)Fröhlich, Dobrynin, Isensee, Gessner, Paxian,
  Pohlmann, Haak, Brune, Früh and Baehr}]{DWD}
Fröhlich, K., Dobrynin, M., Isensee, K., Gessner, C., Paxian, A., Pohlmann,
  H., Haak, H., Brune, S., Früh, B. and Baehr, J. (2020) The german climate
  forecast system: Gcfs.
\newblock \textit{Earth and Space Science Open Archive}, 28.
\newblock \urlprefix\url{https://doi.org/10.1002/essoar.10502582.1}.

\bibitem[{Goldenberg et~al.(2001)Goldenberg, Landsea, Mestas-Nuñez and
  Gray}]{goldberg}
Goldenberg, S.~B., Landsea, C.~W., Mestas-Nuñez, A.~M. and Gray, W.~M. (2001)
  The recent increase in atlantic hurricane activity: Causes and implications.
\newblock \textit{Science}, \textbf{293}, 474--479.

\bibitem[{Gray(1968)}]{gray1968global}
Gray, W.~M. (1968) Global view of the origin of tropical disturbances and
  storms.
\newblock \textit{Monthly Weather Review}, \textbf{96}, 669--700.

\bibitem[{Gray(1984)}]{gray1984atlantic}
--- (1984) Atlantic seasonal hurricane frequency. part ii: Forecasting its
  variability.
\newblock \textit{Monthly Weather Review}, \textbf{112}, 1669--1683.

\bibitem[{Grinsted et~al.(2019)Grinsted, Ditlevsen and
  Christensen}]{Grinsted23942}
Grinsted, A., Ditlevsen, P. and Christensen, J.~H. (2019) Normalized us
  hurricane damage estimates using area of total destruction, 1900-2018.
\newblock \textit{Proceedings of the National Academy of Sciences},
  \textbf{116}, 23942--23946.
\newblock \urlprefix\url{https://www.pnas.org/content/116/48/23942}.

\bibitem[{Gualdi et~al.(2020)Gualdi, Borrelli, Cantelli, Davoli,
  Mar~Chavesmontero, Masina, Navarra, Sanna, Tibaldi et~al.}]{CMCC}
Gualdi, S., Borrelli, A., Cantelli, A., Davoli, G., Mar~Chavesmontero, M.~d.,
  Masina, S., Navarra, A., Sanna, A., Tibaldi, S. et~al. (2020) The new cmcc
  operational seasonal prediction system.
\newblock \textit{CMCC Research Paper}.

\bibitem[{Hersbach et~al.(2020)Hersbach, Bell, Berrisford, Hirahara,
  Hor{\'a}nyi, Mu{\~n}oz-Sabater, Nicolas, Peubey, Radu, Schepers
  et~al.}]{ERA5}
Hersbach, H., Bell, B., Berrisford, P., Hirahara, S., Hor{\'a}nyi, A.,
  Mu{\~n}oz-Sabater, J., Nicolas, J., Peubey, C., Radu, R., Schepers, D. et~al.
  (2020) The era5 global reanalysis.
\newblock \textit{Quarterly Journal of the Royal Meteorological Society},
  \textbf{146}, 1999--2049.

\bibitem[{Hodges(1995)}]{hodges1995feature}
Hodges, K. (1995) Feature tracking on the unit sphere.
\newblock \textit{Monthly Weather Review}, \textbf{123}, 3458--3465.

\bibitem[{Hodges(1996)}]{hodges1996spherical}
--- (1996) Spherical nonparametric estimators applied to the ugamp model
  integration for amip.
\newblock \textit{Monthly Weather Review}, \textbf{124}, 2914--2932.

\bibitem[{Hodges(1999)}]{hodges1999adaptive}
--- (1999) Adaptive constraints for feature tracking.
\newblock \textit{Monthly Weather Review}, \textbf{127}, 1362--1373.

\bibitem[{Hodges et~al.(2017)Hodges, Cobb and Vidale}]{HodgesKevin2017HWAT}
Hodges, K., Cobb, A. and Vidale, P.~L. (2017) How well are tropical cyclones
  represented in reanalysis datasets?
\newblock \textit{Journal of climate}, \textbf{30}, 5243--5264.

\bibitem[{Hodges(1994)}]{hodges1994general}
Hodges, K.~I. (1994) A general method for tracking analysis and its application
  to meteorological data.
\newblock \textit{Monthly Weather Review}, \textbf{122}, 2573--2586.

\bibitem[{Hu et~al.(2018)Hu, Huang and Zhou}]{HU2018640}
Hu, Y., Huang, H. and Zhou, C. (2018) Widening and weakening of the hadley
  circulation under global warming.
\newblock \textit{Science Bulletin}, \textbf{63}, 640--644.
\newblock
  \urlprefix\url{https://www.sciencedirect.com/science/article/pii/S2095927318301919}.

\bibitem[{Jaramillo et~al.(2021)Jaramillo, Dominguez, Raga and
  Quintanar}]{jaramillo2021combined}
Jaramillo, A., Dominguez, C., Raga, G. and Quintanar, A.~I. (2021) The combined
  qbo and enso influence on tropical cyclone activity over the north atlantic
  ocean.
\newblock \textit{Atmosphere}, \textbf{12}, 1588.

\bibitem[{Johnson et~al.(2019)Johnson, Stockdale, Ferranti, Balmaseda, Molteni,
  Magnusson, Tietsche, Decremer, Weisheimer, Balsamo, Keeley, Mogensen, Zuo and
  Monge-Sanz}]{ECMWF}
Johnson, S.~J., Stockdale, T.~N., Ferranti, L., Balmaseda, M.~A., Molteni, F.,
  Magnusson, L., Tietsche, S., Decremer, D., Weisheimer, A., Balsamo, G.,
  Keeley, S. P.~E., Mogensen, K., Zuo, H. and Monge-Sanz, B.~M. (2019) Seas5:
  the new ecmwf seasonal forecast system.
\newblock \textit{Geoscientific Model Development}, \textbf{12}, 1087--1117.
\newblock \urlprefix\url{https://gmd.copernicus.org/articles/12/1087/2019/}.

\bibitem[{Klotzbach et~al.(2019)Klotzbach, Blake, Camp, Caron, Chan, Kang,
  Kuleshov, Lee, Murakami, Saunders, Takaya, Vitart and
  Zhan}]{KlotzbachPhilip2019STCF}
Klotzbach, P., Blake, E., Camp, J., Caron, L.-P., Chan, J.~C., Kang, N.-Y.,
  Kuleshov, Y., Lee, S.-M., Murakami, H., Saunders, M., Takaya, Y., Vitart, F.
  and Zhan, R. (2019) Seasonal tropical cyclone forecasting.
\newblock \textit{Tropical Cyclone Research and Review}, \textbf{8}, 134--149.

\bibitem[{Klotzbach(2011)}]{ImpactonAtlanticBasin}
Klotzbach, P.~J. (2011) El niño–southern oscillation’s impact on atlantic
  basin hurricanes and u.s. landfalls.
\newblock \textit{Journal of Climate}, \textbf{24}, 1252 -- 1263.
\newblock
  \urlprefix\url{https://journals.ametsoc.org/view/journals/clim/24/4/2010jcli3799.1.xml}.

\bibitem[{Klotzbach et~al.(2022)Klotzbach, Wood, Bell, Blake, Bowen, Caron,
  Collins, Gibney, Schreck~III and Truchelut}]{klotzbach2022hyperactive}
Klotzbach, P.~J., Wood, K.~M., Bell, M.~M., Blake, E.~S., Bowen, S.~G., Caron,
  L.-P., Collins, J.~M., Gibney, E.~J., Schreck~III, C.~J. and Truchelut, R.~E.
  (2022) A hyperactive end to the atlantic hurricane season october--november
  2020.
\newblock \textit{Bulletin of the American Meteorological Society},
  \textbf{103}, E110--E128.

\bibitem[{Knapp et~al.(2010)Knapp, Kruk, Levinson, Diamond and
  Neumann}]{IBTrACS}
Knapp, K.~R., Kruk, M.~C., Levinson, D.~H., Diamond, H.~J. and Neumann, C.~J.
  (2010) The international best track archive for climate stewardship
  (ibtracs): Unifying tropical cyclone data.
\newblock \textit{Bulletin of the American Meteorological Society},
  \textbf{91}, 363 -- 376.
\newblock
  \urlprefix\url{https://journals.ametsoc.org/view/journals/bams/91/3/2009bams2755_1.xml}.

\bibitem[{Lim et~al.(2018)Lim, Schubert, Kovach, Molod and
  Pawson}]{lim2018roles}
Lim, Y.-K., Schubert, S.~D., Kovach, R., Molod, A.~M. and Pawson, S. (2018) The
  roles of climate change and climate variability in the 2017 atlantic
  hurricane season.
\newblock \textit{Scientific reports}, \textbf{8}, 1--10.

\bibitem[{Lin et~al.(2020)Lin, Camargo, Patricola, Boucharel, Chand, Klotzbach,
  Chan, Wang, Chang, Li and Jin}]{ENSOcyclonesbook}
Lin, I.-I., Camargo, S.~J., Patricola, C.~M., Boucharel, J., Chand, S.,
  Klotzbach, P., Chan, J. C.~L., Wang, B., Chang, P., Li, T. and Jin, F.-F.
  (2020) \textit{ENSO and Tropical Cyclones}, chap.~17, 377--408.
\newblock American Geophysical Union (AGU).
\newblock
  \urlprefix\url{https://agupubs.onlinelibrary.wiley.com/doi/abs/10.1002/9781119548164.ch17}.

\bibitem[{Lu et~al.(2008)Lu, Chen and Frierson}]{lu2008response}
Lu, J., Chen, G. and Frierson, D.~M. (2008) Response of the zonal mean
  atmospheric circulation to el ni{\~n}o versus global warming.
\newblock \textit{Journal of Climate}, \textbf{21}, 5835--5851.

\bibitem[{MacLachlan et~al.(2015)MacLachlan, Arribas, Peterson, Maidens,
  Fereday, Scaife, Gordon, Vellinga, Williams, Comer et~al.}]{UKMO}
MacLachlan, C., Arribas, A., Peterson, K.~A., Maidens, A., Fereday, D., Scaife,
  A., Gordon, M., Vellinga, M., Williams, A., Comer, R. et~al. (2015) Global
  seasonal forecast system version 5 (glosea5): A high-resolution seasonal
  forecast system.
\newblock \textit{Quarterly Journal of the Royal Meteorological Society},
  \textbf{141}, 1072--1084.

\bibitem[{Vitart et~al.(2007)Vitart, Huddleston, D{\'e}qu{\'e}, Peake, Palmer,
  Stockdale, Davey, Ineson and Weisheimer}]{vitart2007dynamically}
Vitart, F., Huddleston, M.~R., D{\'e}qu{\'e}, M., Peake, D., Palmer, T.~N.,
  Stockdale, T.~N., Davey, M.~K., Ineson, S. and Weisheimer, A. (2007)
  Dynamically-based seasonal forecasts of atlantic tropical storm activity
  issued in june by eurosip.
\newblock \textit{Geophysical Research Letters}, \textbf{34}.

\bibitem[{Vitart and Stockdale(2001)}]{vitart2001seasonal}
Vitart, F. and Stockdale, T.~N. (2001) Seasonal forecasting of tropical storms
  using coupled gcm integrations.
\newblock \textit{Monthly Weather Review}, \textbf{129}, 2521--2537.

\bibitem[{Wang(2004)}]{wang2004enso}
Wang, C. (2004) Enso, atlantic climate variability, and the walker and hadley
  circulations.
\newblock In \textit{The Hadley circulation: present, past and future},
  173--202. Springer.

\bibitem[{Wang et~al.(2014)Wang, Long, Kumar, Wang, Schemm, Zhao, Vecchi,
  Larow, Lim, Schubert, Shaevitz, Camargo, Henderson, Kim, Jonas and
  Walsh}]{climatemodelsenso}
Wang, H., Long, L., Kumar, A., Wang, W., Schemm, J.-K.~E., Zhao, M., Vecchi,
  G.~A., Larow, T.~E., Lim, Y.-K., Schubert, S.~D., Shaevitz, D.~A., Camargo,
  S.~J., Henderson, N., Kim, D., Jonas, J.~A. and Walsh, K. J.~E. (2014) How
  well do global climate models simulate the variability of atlantic tropical
  cyclones associated with enso?
\newblock \textit{Journal of Climate}, \textbf{27}, 5673 -- 5692.
\newblock
  \urlprefix\url{https://journals.ametsoc.org/view/journals/clim/27/15/jcli-d-13-00625.1.xml}.

\bibitem[{Weinkle et~al.(2012)Weinkle, Maue and
  Pielke~Jr}]{weinkle2012historical}
Weinkle, J., Maue, R. and Pielke~Jr, R. (2012) Historical global tropical
  cyclone landfalls.
\newblock \textit{Journal of Climate}, \textbf{25}, 4729--4735.

\bibitem[{Willoughby(1999)}]{willoughby1999hurricane}
Willoughby, H. (1999) Hurricane heat engines.
\newblock \textit{Nature}, \textbf{401}, 649--650.

\bibitem[{Xie et~al.(2005)Xie, Yan, Pietrafesa, Morrison and
  Karl}]{xie2005climatology}
Xie, L., Yan, T., Pietrafesa, L.~J., Morrison, J.~M. and Karl, T. (2005)
  Climatology and interannual variability of north atlantic hurricane tracks.
\newblock \textit{Journal of climate}, \textbf{18}, 5370--5381.

\bibitem[{Yang et~al.(2018)Yang, Li, Yu, Hu, Dong and He}]{YangSong2018ENOa}
Yang, S., Li, Z., Yu, J.-Y., Hu, X., Dong, W. and He, S. (2018) El
  niño–southern oscillation and its impact in the changing climate.
\newblock \textit{National science review}, \textbf{5}, 840--857.

\end{thebibliography}

\newpage
\section{Figures}

\newpage 

\newpage 
\begin{figure}[!htb]
    \centering
    \includegraphics[width= 0.8\textwidth]{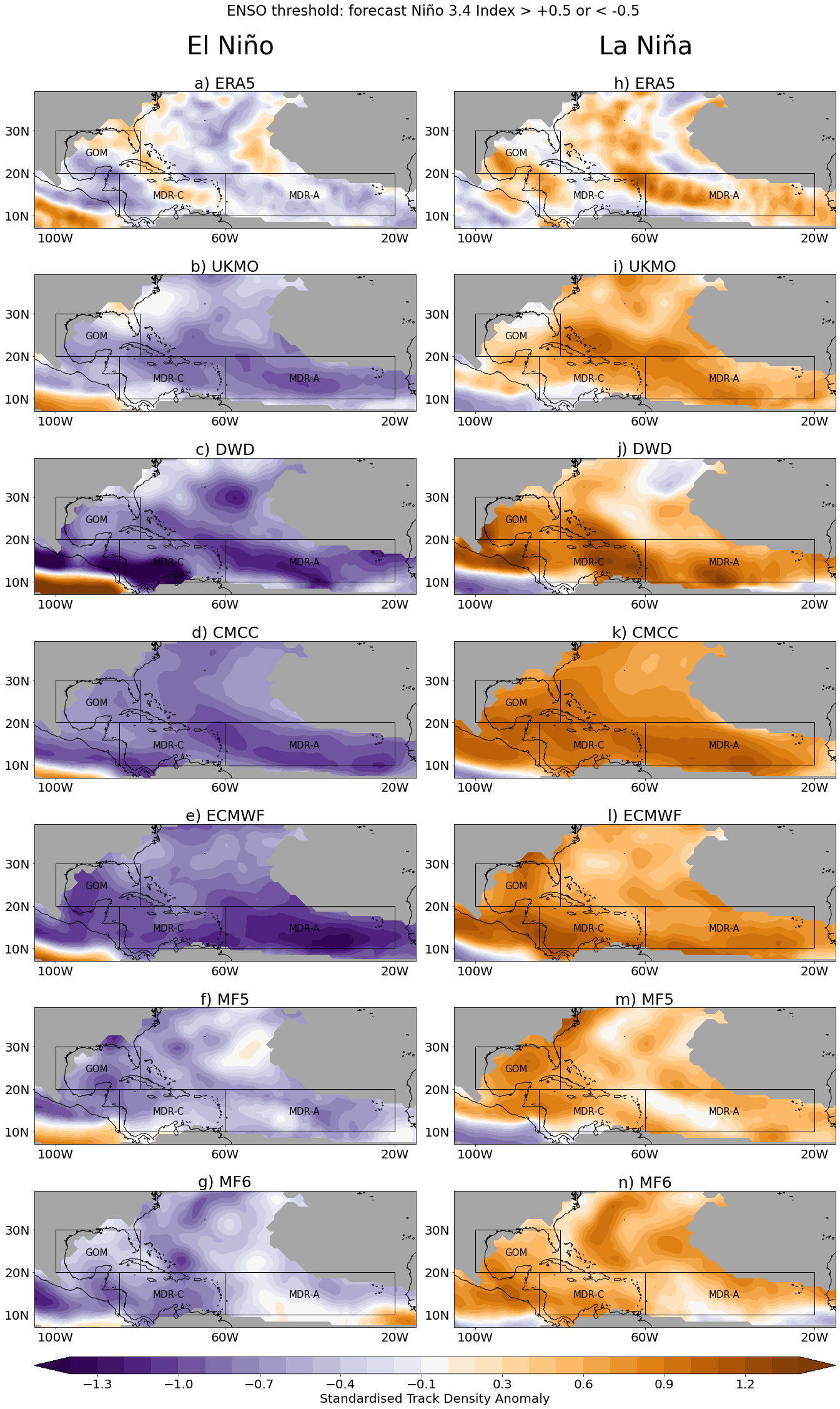}
    \caption{ENSO track composites for El Niño (a-g) \& La Niña (h-n). Map of standardised track density anomalies during El Niño \& La Niña years predicted by the respective model. Panels a) and h) for reanalysis data and all other panels for seasonal forecast models. Areas are masked where there is less than half the basin average TC track density in ERA5.}
    \label{composites}
\end{figure}

\newpage 
\begin{figure}[!htb]
    \centering
    \includegraphics[width = 0.4\textwidth]{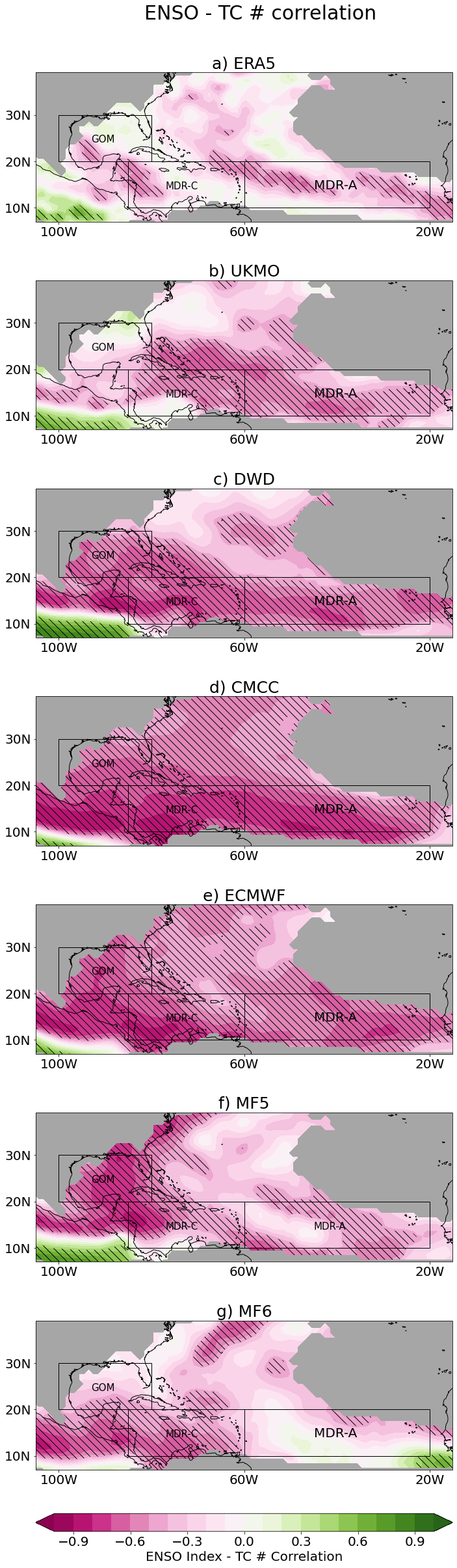}
    \caption{ENSO-TC correlation map. Correlation coefficient of interannual variability between NINO 3.4 SST index and number of TCs in the North Atlantic basin for JASO from 1993-2014. Panel a) for ERA5 data and panels b)-g) for seasonal forecast models. Areas are masked where there is less than half the basin average TC track density in ERA5. Shaded areas correspond to correlations significant at the 5\% level.}
    \label{TC-ENSO map}
\end{figure}

\newpage 
\begin{figure}[!htb]
    \centering
    \includegraphics[width = 0.65\textwidth]{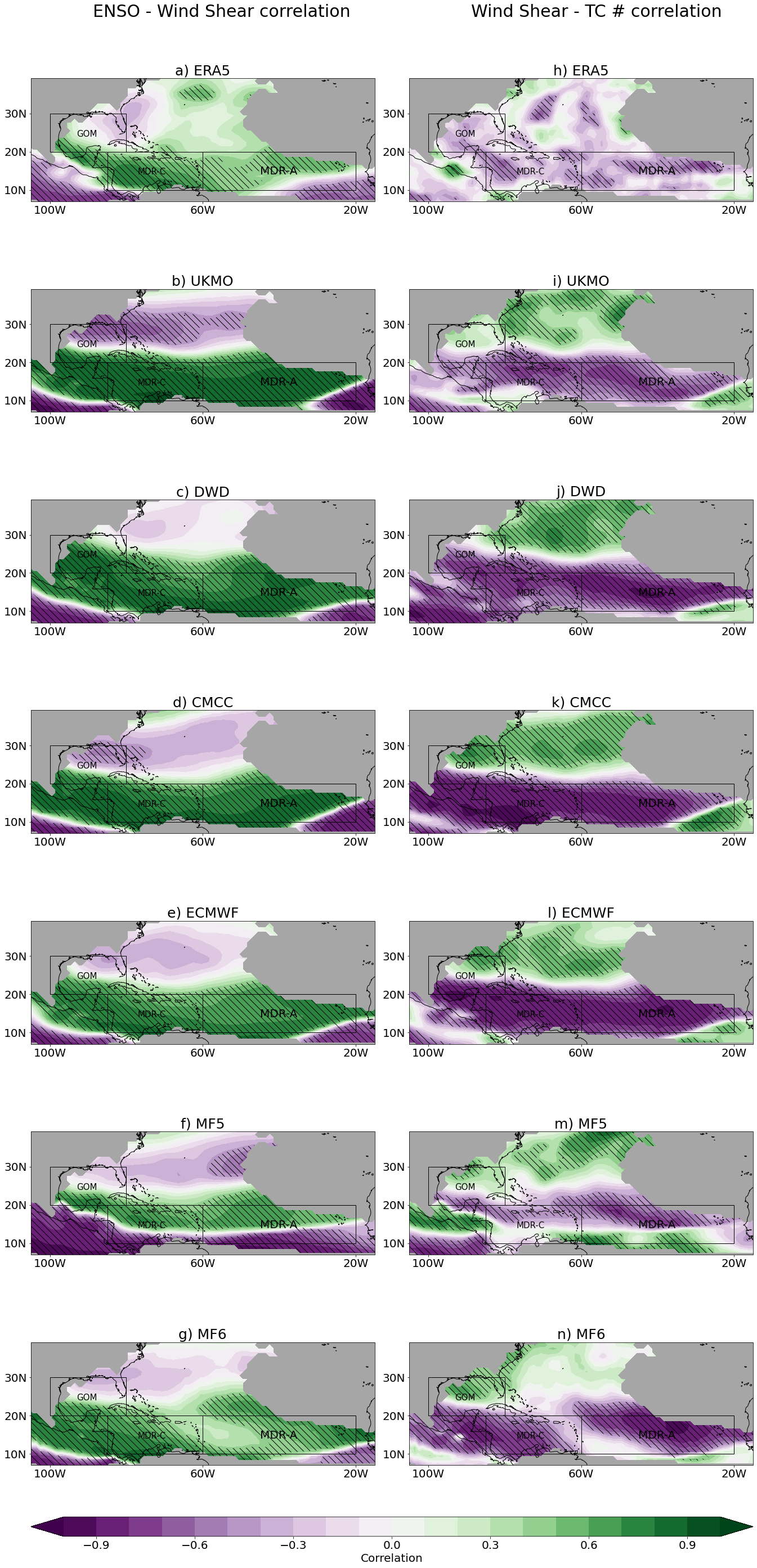}
    \caption{ENSO - Wind Shear \& Wind Shear - TC correlations. Left column: Correlation coefficients of interannual variability between NINO 3.4 SST index and wind shear in the North Atlantic basin for JASO from 1993-2014. Right column: Correlation coefficients of interannual variability between wind shear and total number of TCs in the North Atlantic basin for the same periods. Panels a) and h) for reanalysis data and all other panels for seasonal forecast models. Areas are masked where there is less than half the basin average TC track density in ERA5. Shaded areas correspond to correlations significant at the 5\% level.}
    \label{Shear map}
\end{figure}

\clearpage
\newpage
\begin{center}
\textbf{\large Supplementary Material}
\end{center}
\setcounter{equation}{0}
\setcounter{figure}{0}
\setcounter{table}{0}
\setcounter{section}{0}

\renewcommand{\thetable}{S\arabic{table}}
\setcounter{page}{1}
\makeatletter

\renewcommand{\theequation}{S\arabic{equation}}
\renewcommand{\thefigure}{S\arabic{figure}}
\renewcommand{\bibnumfmt}[1]{[S#1]}
\renewcommand{\citenumfont}[1]{S#1}

\section{Model Biases}

We conducted an analysis of forecast model biases and found significant effects. The models have considerably different average SSTs in the Niño 3.4 region, although they all have fairly similar levels of variability. We conclude that many of the forecast models have cold biases. They also predict very different numbers of storms, ranging from an average of $2.9 \pm 0.6$ per year (DWD) to $18.1 \pm 2.4$ (UKMO), compared to an ERA5 count of $14.5 \pm 4.4$. However there appears to be no correspondence between the average SST and average number of storms formed. The variation in storm numbers is lower in the forecasts than best track data. 

\bigbreak
\noindent

\begin{figure}[!htb]
    \centering
    \includegraphics[width=\textwidth]{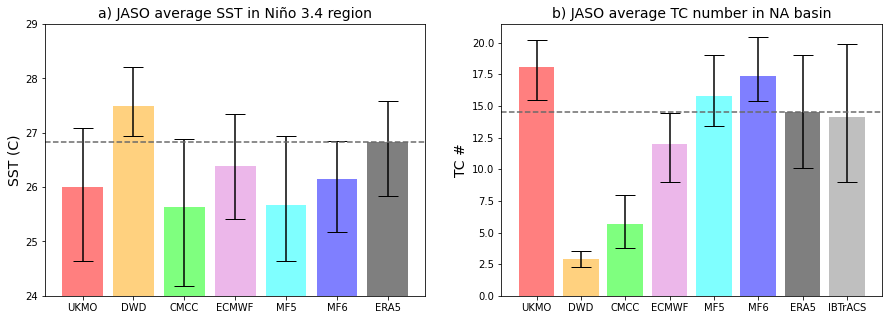}
    \caption{a) shows the JASO average SST in the Niño 3.4 region, b) shows the JASO average TC number in the North Atlantic. Error bars in both plots correspond to values between the 10th \& 90th percentiles of the distribution of SST indices/TC number over the 1993-2014 period.}
    \label{SST,TC bias}
\end{figure}

\section{Model Data}

\begin{table}[ht]
\centering
\captionsetup{justification=centering}
\caption{Seasonal forecast datasets used in this study}
\label{model table}
\begin{adjustbox}{width={0.7\textwidth}}
\begin{tabular}{cccc}
\hline
Model & Atmosphere resolution & Ocean resolution & Ensemble size\\
\hline
UKMO & N216/L85 $\approx$ 90 km & 0.25\textdegree/L75 & 28\\
DWD & T127/L91 $\approx$ 100 km & 0.4\textdegree/L70 & 30\\
CMCC & $\approx$ 110 km/L46 & 0.25\textdegree/L50 & 40\\
ECMWF & TCO319/L91 $\approx$ 32 km & 0.25\textdegree/L75 & 25\\
MF5 & TL255/L91 $\approx$ 80 km & 1\textdegree/42 levels & 15\\
MF6 & TL359/L91 $\approx$ 50 km & 1\textdegree/75 levels & 25\\

\hline 
\end{tabular}
\end{adjustbox}
\end{table}

\newpage

\section{TC-ENSO Composite Analysis}

\begin{figure}[!htb]
    \centering
    \includegraphics[width=0.8\textwidth]{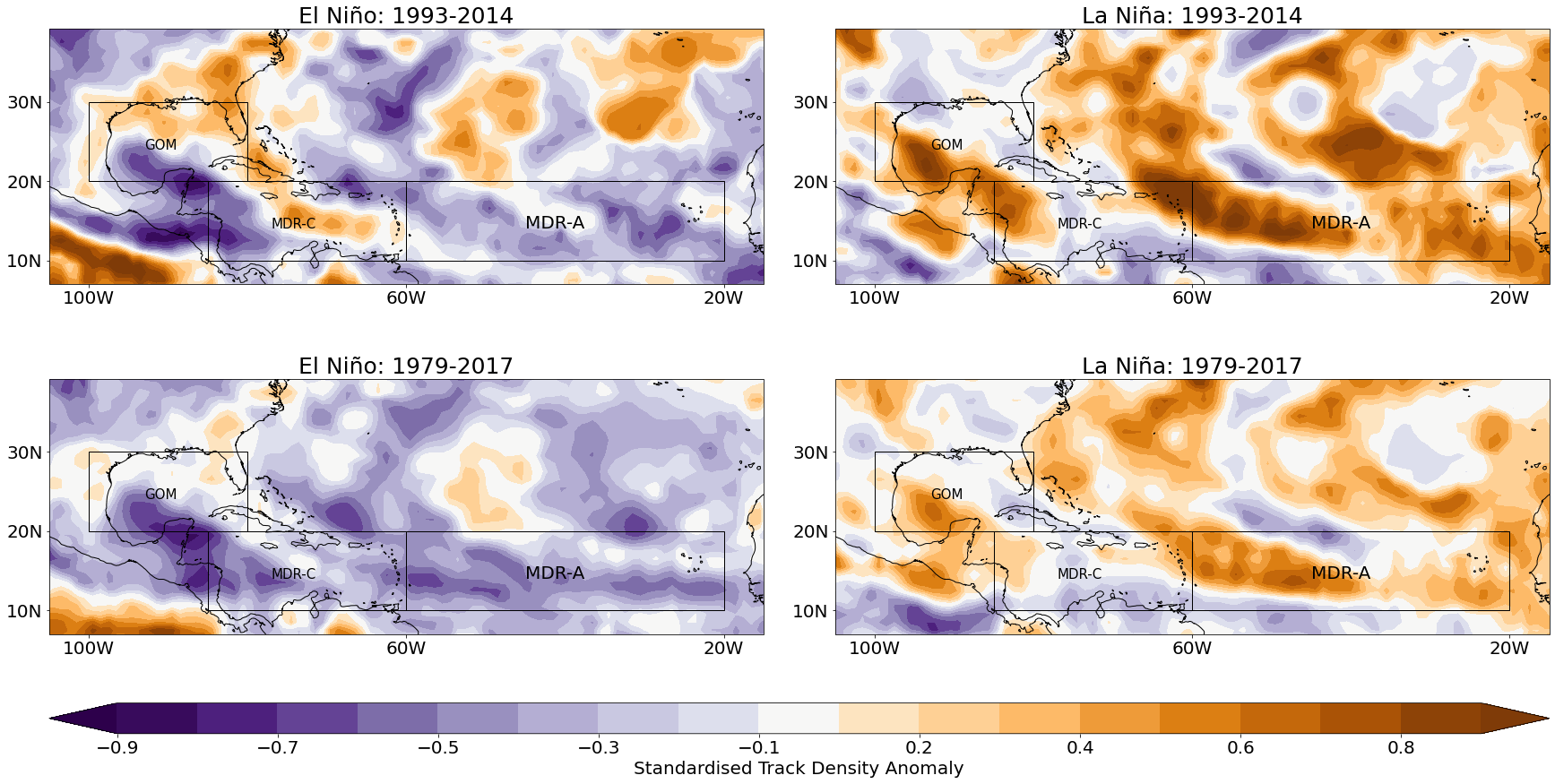}
    \caption{ERA5 standardised track density anomaly composites. The columns shows the anomalies for both ENSO phases, and the rows show how the pattern changes depending on the years studied.}
    \label{era5 comparison}
\end{figure}

\begin{table}[!htb]
\captionsetup{justification=centering}
\caption{Years used for composite analysis}
\label{years table}
\begin{adjustbox}{width={\textwidth}}
\begin{tabular}{lll}
\hline
Model & El Niño Years & La Niña Years\\
\hline
UKMO & 1993, 1994, 1997, 2002, 2006, 2009, 2012, 2014 & 1995, 1996, 1998, 1999, 2003, 2007, 2010 \\ 
DWD & 1993, 1997, 2014 & 1996, 1999, 2000, 2007 \\ 
CMCC & 1997, 2002, 2006, 2009, 2012, 2014 & 1995, 1996, 1998, 1999, 2003, 2007, 2010 \\ 
ECMWF & 1997, 2002, 2009, 2012, 2014 & 1996, 1998, 1999, 2003, 2007, 2010 \\ 
MF5 & 1997, 2002, 2006, 2009, 2012, 2014 & 1995, 1996, 1998, 1999, 2000, 2003, 2005, 2007, 2010 \\ 
MF6 & 1993, 1997, 2002, 2006, 2009, 2012 & 1998, 1999, 2000, 2003, 2007, 2010 \\ 
ERA5/IBTrACS & 1997, 2002, 2004, 2006, 2009, 2012 & 1995, 1998, 1999, 2007, 2010, 2011 \\ 
\hline 
\end{tabular}
\end{adjustbox}
\end{table}

\section{Humidity Correlation Analysis}

\begin{figure}
    \centering
    \includegraphics[width = 0.54\textwidth]{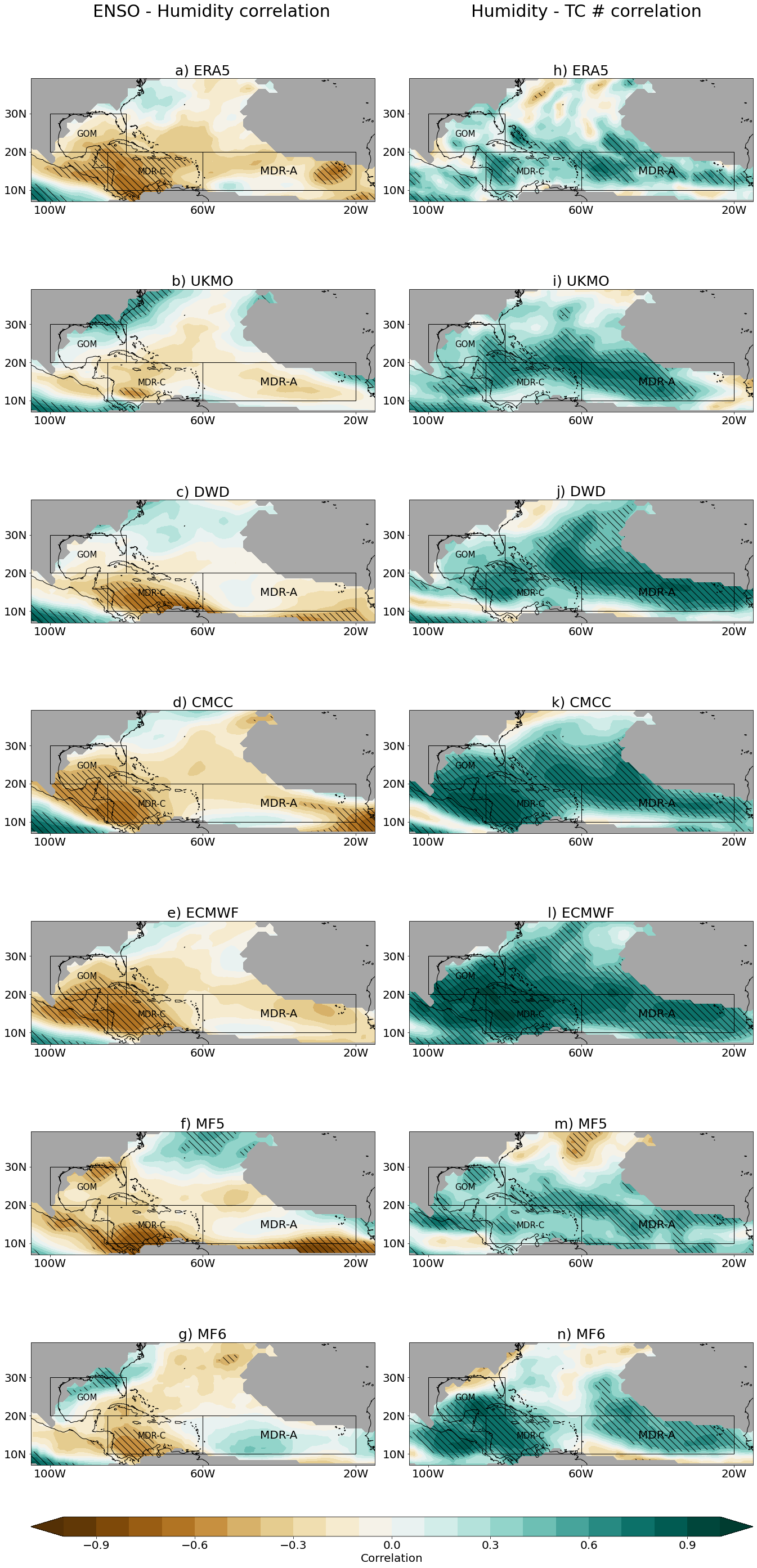}
    \caption{ENSO - Humidity \& Wind Shear - Humidity correlations. Left column: Correlation coefficients of interannual variability between NINO 3.4 SST index and 500hPa specific humidity in the North Atlantic basin for JASO from 1993-2014. Right column: Correlation coefficients of interannual variability between 500hPa specific humidity and total number of TCs in the North Atlantic basin for the same periods. Panels a) and h) for reanalysis data and all other panels for seasonal forecast models. Areas are masked where there is less than half the basin average TC track density in ERA5. Shaded areas correspond to correlations significant at the 5\% level.}
    \label{Humidity map}
\end{figure}

In Figure \ref{Humidity map} ERA5 shows a negative correlation between humidity and ENSO across all of the extended tropical NA basin, with a large area of statistically significant negative correlations in the MDR-C. The USA coastline near Florida shows no clear signal. All models capture the correlation across the tropics, and most have an area of significance in MDR-C. However around Florida some models show stronger correlations, which is not replicated in ERA5 and is possibly related to errors in this region shown in Figs. \ref{composites} \& \ref{TC-ENSO map}. For the relationship between humidity and TC numbers, ERA5 shows a large area of strong positive correlation across the tropics, and no correlation in the subtropics. Generally, the models simulate this pattern, although in most it is too strong and widespread.

\newpage 

\end{document}